\RequirePackage{lineno}

 \documentclass[aps,prd,twocolumn,superscriptaddress,showpacs]{revtex4}

\usepackage{graphicx}
\usepackage{dcolumn}
\usepackage{bm}
\usepackage{color}

\newcommand{\ppbar}{$\bar{p}p$}                                                                                             %

\newcommand{\dedx}{\ensuremath{\mathit{dE/dx}}}                                 

\newcommand{\sigLxy}{\ensuremath{L_{xy}(B)/\sigma_{L_{xy}(B)}}}

\newcommand{\babar}{{\emph {B\scriptsize{A}\small B\scriptsize{AR}}}}

\begin{document}

 \title{Measurements of branching fraction ratios and $\mathit{CP}$-asymmetries in suppressed $B^{-} \to D(\to K^+\pi^-)K^{-}$ and $B^{-} \to D(\to K^+\pi^-)\pi^{-}$ decays}
\affiliation{Institute of Physics, Academia Sinica, Taipei, Taiwan 11529, Republic of China} 
\affiliation{Argonne National Laboratory, Argonne, Illinois 60439, USA} 
\affiliation{University of Athens, 157 71 Athens, Greece} 
\affiliation{Institut de Fisica d'Altes Energies, ICREA, Universitat Autonoma de Barcelona, E-08193, Bellaterra (Barcelona), Spain} 
\affiliation{Baylor University, Waco, Texas 76798, USA} 
\affiliation{Istituto Nazionale di Fisica Nucleare Bologna, $^{aa}$University of Bologna, I-40127 Bologna, Italy} 
\affiliation{University of California, Davis, Davis, California 95616, USA} 
\affiliation{University of California, Los Angeles, Los Angeles, California 90024, USA} 
\affiliation{Instituto de Fisica de Cantabria, CSIC-University of Cantabria, 39005 Santander, Spain} 
\affiliation{Carnegie Mellon University, Pittsburgh, Pennsylvania 15213, USA} 
\affiliation{Enrico Fermi Institute, University of Chicago, Chicago, Illinois 60637, USA}
\affiliation{Comenius University, 842 48 Bratislava, Slovakia; Institute of Experimental Physics, 040 01 Kosice, Slovakia} 
\affiliation{Joint Institute for Nuclear Research, RU-141980 Dubna, Russia} 
\affiliation{Duke University, Durham, North Carolina 27708, USA} 
\affiliation{Fermi National Accelerator Laboratory, Batavia, Illinois 60510, USA} 
\affiliation{University of Florida, Gainesville, Florida 32611, USA} 
\affiliation{Laboratori Nazionali di Frascati, Istituto Nazionale di Fisica Nucleare, I-00044 Frascati, Italy} 
\affiliation{University of Geneva, CH-1211 Geneva 4, Switzerland} 
\affiliation{Glasgow University, Glasgow G12 8QQ, United Kingdom} 
\affiliation{Harvard University, Cambridge, Massachusetts 02138, USA} 
\affiliation{Division of High Energy Physics, Department of Physics, University of Helsinki and Helsinki Institute of Physics, FIN-00014, Helsinki, Finland} 
\affiliation{University of Illinois, Urbana, Illinois 61801, USA} 
\affiliation{The Johns Hopkins University, Baltimore, Maryland 21218, USA} 
\affiliation{Institut f\"{u}r Experimentelle Kernphysik, Karlsruhe Institute of Technology, D-76131 Karlsruhe, Germany} 
\affiliation{Center for High Energy Physics: Kyungpook National University, Daegu 702-701, Korea; Seoul National University, Seoul 151-742, Korea; Sungkyunkwan University, Suwon 440-746, Korea; Korea Institute of Science and Technology Information, Daejeon 305-806, Korea; Chonnam National University, Gwangju 500-757, Korea; Chonbuk National University, Jeonju 561-756, Korea} 
\affiliation{Ernest Orlando Lawrence Berkeley National Laboratory, Berkeley, California 94720, USA} 
\affiliation{University of Liverpool, Liverpool L69 7ZE, United Kingdom} 
\affiliation{University College London, London WC1E 6BT, United Kingdom} 
\affiliation{Centro de Investigaciones Energeticas Medioambientales y Tecnologicas, E-28040 Madrid, Spain} 
\affiliation{Massachusetts Institute of Technology, Cambridge, Massachusetts 02139, USA} 
\affiliation{Institute of Particle Physics: McGill University, Montr\'{e}al, Qu\'{e}bec, Canada H3A~2T8; Simon Fraser University, Burnaby, British Columbia, Canada V5A~1S6; University of Toronto, Toronto, Ontario, Canada M5S~1A7; and TRIUMF, Vancouver, British Columbia, Canada V6T~2A3} 
\affiliation{University of Michigan, Ann Arbor, Michigan 48109, USA} 
\affiliation{Michigan State University, East Lansing, Michigan 48824, USA}
\affiliation{Institution for Theoretical and Experimental Physics, ITEP, Moscow 117259, Russia}
\affiliation{University of New Mexico, Albuquerque, New Mexico 87131, USA} 
\affiliation{Northwestern University, Evanston, Illinois 60208, USA} 
\affiliation{The Ohio State University, Columbus, Ohio 43210, USA} 
\affiliation{Okayama University, Okayama 700-8530, Japan} 
\affiliation{Osaka City University, Osaka 588, Japan} 
\affiliation{University of Oxford, Oxford OX1 3RH, United Kingdom} 
\affiliation{Istituto Nazionale di Fisica Nucleare, Sezione di Padova-Trento, $^{bb}$University of Padova, I-35131 Padova, Italy} 
\affiliation{LPNHE, Universite Pierre et Marie Curie/IN2P3-CNRS, UMR7585, Paris, F-75252 France} 
\affiliation{University of Pennsylvania, Philadelphia, Pennsylvania 19104, USA}
\affiliation{Istituto Nazionale di Fisica Nucleare Pisa, $^{cc}$University of Pisa, $^{dd}$University of Siena and $^{ee}$Scuola Normale Superiore, I-56127 Pisa, Italy} 
\affiliation{University of Pittsburgh, Pittsburgh, Pennsylvania 15260, USA} 
\affiliation{Purdue University, West Lafayette, Indiana 47907, USA} 
\affiliation{University of Rochester, Rochester, New York 14627, USA} 
\affiliation{The Rockefeller University, New York, New York 10065, USA} 
\affiliation{Istituto Nazionale di Fisica Nucleare, Sezione di Roma 1, $^{ff}$Sapienza Universit\`{a} di Roma, I-00185 Roma, Italy} 

\affiliation{Rutgers University, Piscataway, New Jersey 08855, USA} 
\affiliation{Texas A\&M University, College Station, Texas 77843, USA} 
\affiliation{Istituto Nazionale di Fisica Nucleare Trieste/Udine, I-34100 Trieste, $^{gg}$University of Udine, I-33100 Udine, Italy} 
\affiliation{University of Tsukuba, Tsukuba, Ibaraki 305, Japan} 
\affiliation{Tufts University, Medford, Massachusetts 02155, USA} 
\affiliation{University of Virginia, Charlottesville, Virginia 22906, USA}
\affiliation{Waseda University, Tokyo 169, Japan} 
\affiliation{Wayne State University, Detroit, Michigan 48201, USA} 
\affiliation{University of Wisconsin, Madison, Wisconsin 53706, USA} 
\affiliation{Yale University, New Haven, Connecticut 06520, USA} 
\author{T.~Aaltonen}
\affiliation{Division of High Energy Physics, Department of Physics, University of Helsinki and Helsinki Institute of Physics, FIN-00014, Helsinki, Finland}
\author{B.~\'{A}lvarez~Gonz\'{a}lez$^w$}
\affiliation{Instituto de Fisica de Cantabria, CSIC-University of Cantabria, 39005 Santander, Spain}
\author{S.~Amerio}
\affiliation{Istituto Nazionale di Fisica Nucleare, Sezione di Padova-Trento, $^{bb}$University of Padova, I-35131 Padova, Italy} 

\author{D.~Amidei}
\affiliation{University of Michigan, Ann Arbor, Michigan 48109, USA}
\author{A.~Anastassov}
\affiliation{Northwestern University, Evanston, Illinois 60208, USA}
\author{A.~Annovi}
\affiliation{Laboratori Nazionali di Frascati, Istituto Nazionale di Fisica Nucleare, I-00044 Frascati, Italy}
\author{J.~Antos}
\affiliation{Comenius University, 842 48 Bratislava, Slovakia; Institute of Experimental Physics, 040 01 Kosice, Slovakia}
\author{G.~Apollinari}
\affiliation{Fermi National Accelerator Laboratory, Batavia, Illinois 60510, USA}
\author{J.A.~Appel}
\affiliation{Fermi National Accelerator Laboratory, Batavia, Illinois 60510, USA}
\author{A.~Apresyan}
\affiliation{Purdue University, West Lafayette, Indiana 47907, USA}
% ...
\collaboration{CDF Collaboration\footnote{With visitors from $^a$Istituto Nazionale di Fisica Nucleare, Sezione di Cagliari, 09042 Monserrato (Cagliari), Italy,
$^b$University of CA Irvine, Irvine, CA  92697, USA,
$^c$University of CA Santa Barbara, Santa Barbara, CA 93106, USA,
$^d$University of CA Santa Cruz, Santa Cruz, CA  95064, USA,
$^e$CERN,CH-1211 Geneva, Switzerland,
$^f$Cornell University, Ithaca, NY  14853, USA, 
$^g$University of Cyprus, Nicosia CY-1678, Cyprus, 
$^h$Office of Science, U.S. Department of Energy, Washington, DC 20585, USA,
$^i$University College Dublin, Dublin 4, Ireland,
$^j$University of Fukui, Fukui City, Fukui Prefecture, Japan 910-0017,
$^k$Universidad Iberoamericana, Mexico D.F., Mexico,
$^l$Iowa State University, Ames, IA  50011, USA,
$^m$University of Iowa, Iowa City, IA  52242, USA,
$^n$Kinki University, Higashi-Osaka City, Japan 577-8502,
$^o$Kansas State University, Manhattan, KS 66506, USA,
$^p$University of Manchester, Manchester M13 9PL, United Kingdom,
$^q$Queen Mary, University of London, London, E1 4NS, United Kingdom,
$^r$University of Melbourne, Victoria 3010, Australia,
$^s$Muons, Inc., Batavia, IL 60510, USA,
$^t$Nagasaki Institute of Applied Science, Nagasaki, Japan, 
$^u$National Research Nuclear University, Moscow, Russia,
$^v$University of Notre Dame, Notre Dame, IN 46556, USA,
$^w$Universidad de Oviedo, E-33007 Oviedo, Spain, 
$^x$Texas Tech University, Lubbock, TX  79609, USA,
$^y$Universidad Tecnica Federico Santa Maria, 110v Valparaiso, Chile,
$^z$Yarmouk University, Irbid 211-63, Jordan,
$^{hh}$On leave from J.~Stefan Institute, Ljubljana, Slovenia, 
}}
\noaffiliation

 
  \date{\today}
  
  \begin{abstract}
  
We report the first reconstruction in hadron collisions of the suppressed decays $B^{-} \to D(\to K^+\pi^-)K^{-}$ and $B^{-} \to D(\to K^+\pi^-)\pi^{-}$, sensitive to the CKM phase $\gamma$, using data from 7 fb$^{-1}$ of integrated luminosity collected by the CDF II detector at the Tevatron collider.  
 We reconstruct a signal for the $B^{-} \to D(\to K^+\pi^-)K^{-}$ suppressed mode with a significance of 3.2 standard deviations, and measure the ratios of the suppressed to favored branching fractions 
 $R(K) = [22.0\pm 8.6(\rm{stat})\pm 2.6(\rm{syst})]\times 10^{-3}$, 
$R^+ (K)   =  [42.6\pm 13.7(\rm{stat})\pm 2.8(\rm{syst})]\times 10^{-3}$,
$R^- (K)   =  [3.8\pm 10.3(\rm{stat})\pm 2.7(\rm{syst})]\times 10^{-3}$
  as well as the direct  $\mathit{CP}$-violating asymmetry
  $A(K) = -0.82\pm 0.44(\rm{stat})\pm 0.09(\rm{syst})$ of this mode. 
   Corresponding quantities for $B^{-} \to D(\to K^+\pi^-)\pi^{-}$ decay are also reported.

\end{abstract}
  
  
\pacs{13.25.Hw 11.30.Er 14.40.Nd}  
  
  
\maketitle
  
  
  
The measurement of $\mathit{CP}$-violating asymmetries and branching ratios of $B^{-} \rightarrow D K^{-}$~\cite{charge_conjugate} decay modes allows a theoretically clean extraction of the phase $\gamma = \arg(-V_{ud}V^*_{ub}/V_{cd}V^*_{cb})$ of the Cabibbo-Kobayashi-Maskawa quark-mixing matrix $V_{\it CKM}$, a fundamental parameter of the standard model~\cite{Kobayashi}.  
In these decays the interference between the first order tree amplitudes of the $b\to c\bar{u}s$ and $b\to u\bar{c}s$ processes leads to observables that depend on their relative weak phase $\gamma$, their relative strong phase $\delta_{B}$, and the magnitude of the amplitude ratio $r_{B}$~\cite{rb_def}.
These quantities can all be extracted from data by combining several experimental observables. This can be achieved in several ways, using a variety of $D$ decay channels~\cite{Gronau, Atwood, Giri}. 
An accurate knowledge of the value of $\gamma$ is instrumental in establishing  the possible presence of additional non-standard model $\mathit{CP}$-violating phases in processes where higher-order diagrams are involved~\cite{fleischer,dunietz}.
Its current determination has a relative uncertainty, dominated by statistical uncertainties, between $15$ and  $20\%$, depending on the method~\cite{hfag}. 

%

A promising class of processes consists of $B$ meson decays that are a coherent superposition of the
color favored $B^- \rightarrow D^0 K^-$ followed by the doubly Cabibbo suppressed decay 
$D^0\to K^+\pi^-$, and of the color suppressed $B^- \rightarrow \bar{D}^0 K^-$ followed by 
the Cabibbo favored decay $\bar{D}^0 \to K^+ \pi^-$. 
The magnitude of the two amplitudes is comparable, 
allowing for large $CP$-violating asymmetries sensitive to the phase $\gamma$. 
The following observables can be defined~\cite{Atwood}:
{\footnotesize \begin{eqnarray}
 R(K)  & = &  \frac{\mathcal{B}(B^-\rightarrow [K^+ \pi^-]_{D}K^-)+\mathcal{B}(B^+\rightarrow [K^-\pi^+]_{D}K^+)}{\mathcal{B}(B^-\rightarrow [K^- \pi^+]_{D}K^-)+\mathcal{B}(B^+\rightarrow [K^+\pi^-]_{D}K^+)}, \nonumber \\
R^{\pm}(K) & = & \frac{\mathcal{B}(B^{\pm} \rightarrow [K^{\mp} \pi^{\pm}]_{D} K^{\pm})}{\mathcal{B}(B^{\pm} \rightarrow [K^{\pm} \pi^{\mp}]_{D} K^{\pm})},\nonumber \\
A(K)  & = &  \frac{\mathcal{B}(B^-\rightarrow [K^+\pi^-]_{D}K^-)-\mathcal{B}(B^+\rightarrow [K^-\pi^+]_{D}K^+)}{\mathcal{B}(B^-\rightarrow [K^+\pi^-]_{D}K^-)+\mathcal{B}(B^+\rightarrow [K^-\pi^+]_{D}K^+)},\nonumber
\end{eqnarray}}where $B^-\rightarrow [K^+ \pi^-]_{D}K^-$ is the suppressed ($\mathit{sup}$) mode and $B^-\rightarrow [K^- \pi^+]_{D}K^-$ is the favored ($\mathit{fav}$) mode.
In the approximation of negligible $\mathit{CP}$-violation in $D$ decays and negligible $D^0-\bar{D}^0$ mixing, whose effects were shown to be small in Ref.~\cite{d0mixing}, these quantities are related to the CKM phase $\gamma$ by the equations~\cite{Atwood}  
$R  =  r_D^2 + r_B^2 + 2r_Dr_B \cos{\gamma}\cos{(\delta_B+\delta_D)}$, 
$R^{\pm}  =  r_D^2 + r_B^2 + 2r_Dr_B\cos{(\delta_B+\delta_D \pm \gamma)}$, 
and 
$A  =  2r_Br_D\sin{\gamma}\sin{(\delta_B+\delta_D)}/R$, where $r_D = \left | \frac{A(D^0\rightarrow K^+\pi^-)}{A(D^0 \rightarrow K^-\pi^+)}\right|$ and $\delta_D$ is the corresponding relative strong phase. 
The smallness of the product of branching fractions for these suppressed final states ($\mathcal{O}(10^{-7})$) has been a strong limitation to their use in $\gamma$ determinations. Evidence for the suppressed $B^-\to DK^-$ channel has only recently been obtained by the Belle collaboration~\cite{ckm_belle}. 
The large production rate of $B$ mesons available at hadron colliders offers a unique opportunity for improving the experimental determination of the angle $\gamma$.  Measurements of branching fractions and $\mathit{CP}$-violating asymmetries of $B^{-} \rightarrow D K^{-}$ modes in less suppressed final states of the $D$ meson ($\mathit{CP}$-even modes $K^- K^+$ and $\pi^- \pi^+$) have already been performed in hadron collisions~\cite{prd_glw}. However, the small decay rates along with large potential backgrounds from misidentified favored decays, which only differ for the identity of the final particles, make the reconstruction of suppressed modes in hadron collisions significantly more challenging.

In this Letter, we describe the first reconstruction of  $B^-\to D_{\mathit{sup}}K^-$ modes performed in hadron collisions,  
based on data from a total integrated luminosity of 7 fb$^{-1}$ of \ppbar\ collisions at $\sqrt{s} = 1.96$ TeV, collected by the upgraded Collider Detector (CDF II) at the Fermilab Tevatron. 
We report measurements of  $R(K)$, $R^{\pm}(K)$,  and $A(K)$ for those modes. We also report measurements related to the corresponding $D\pi^-$ modes, since measurable, albeit smaller, $\gamma$-dependent asymmetries may also be found in these modes~\cite{hfag}. The maximum possible value of the asymmetry is $A_{\rm max} = 2r_B r_D/(r^2_B + r^2_D)$, where $r_B$ can be $r_B(K)$ or $r_B(\pi)$. Taking into account the CKM structure of the contributing processes, we expect that $r_B(\pi)$ is suppressed by a factor $|V_{cd}V_{us}/V_{ud}V_{cs}| \sim \tan^2 \theta_C$ with respect to $r_B(K)$, where $\theta_C$ is the Cabibbo angle, and we assume the same color suppression factor for both $DK$ and $D\pi$ modes.
Using $r_B(K) = 0.103 ^{+0.015}_{-0.024}$~\cite{hfag}, $r_B(\pi) \sim 0.005$~\cite{hfag}, and $r^2_D = (3.80 \pm 0.18)\times 10^{-3}$~\cite{pdg}, we expect $A_{\rm max}(K) \approx 0.90 $ and $A_{\rm max}(\pi) \approx 0.16$.

  CDF II is a multipurpose magnetic spectrometer surrounded by
  calorimeters and muon detectors, and is described in detail elsewhere~\cite{Acosta:2004yw,CDF-coordinates}.
 The resolution on transverse momentum of charged particles is $\sigma_{p_{T}}/p_{T} \simeq 
  0.07\%\, p_{T}$/(GeV/$c$), corresponding to a typical mass resolution of 18 MeV/c$^{2}$ for our signals.
  The specific ionization energy loss \dedx\ of charged particles can be measured from the charge collected by a gaseous drift chamber (COT), and provides 1.5$\sigma$ separation between pion and kaon particles for $p> 2$~GeV/$c$.
 %
 %
  Candidate events for this analysis are selected by a three-level on-line event-selection system (trigger).  At level 1, charged particles are reconstructed in the
  COT by a hardware processor, the extremely fast
  tracker XFT~\cite{Thomson:2002xp}. Two
  oppositely charged particles are required, with transverse momenta $p_T \geq 2$
  GeV/$c$ and scalar sum $p_{T1}+p_{T2}\geq 5.5$ GeV/$c$.  At level 2,
  another processor, a silicon vertex trigger (SVT)~\cite{Ashmanskas:2003gf}, associates $r-\phi$ position measurements from an inner silicon detector with XFT tracks.  This provides a precise 
  measurement of the track impact parameter $d_0$, the transverse distance of closest
  approach to the beam line.  The resolution
  of the impact parameter measurement is 50 $\mu$m for particles with 
  $p_T$ of about 2 GeV/$c$, including a $\approx 30$~$\mu$m 
  contribution due to the transverse beam size, and improves for 
  higher transverse momenta. 
  
  We select $B$ hadron candidates by requiring two SVT tracks with 120
  $\leq d_0 \leq$ 1000 $\mu$m.  To reduce background from light-quark
  jet pairs, the two trigger tracks are required to have an opening angle in the transverse plane $2^{\circ} \leq \Delta\phi \leq
  90^{\circ}$, and to satisfy the requirement $L_{xy} > 200$ $\mu$m,
  where $L_{xy}$ is defined as the distance in the transverse plane from the beam line to the reconstructed two-track vertex. The level 1 and 2 trigger requirements are then confirmed at trigger level 3, where the event is fully reconstructed in software.
  
The events collected by the trigger are further selected by searching for a pair of oppositely charged particles compatible with a two-body $D$ decay.
The invariant mass $M_{D}$ of the pair is reconstructed for both pion and kaon assignments of particle identities. Events are accepted for the analysis only when one of the possible masses is compatible with the nominal $D$ mass $1.8495\le M_D\le  1.8815$ GeV$/c^2$, and the alternative combination, $M_{\it SW}(D)$, 
is outside a veto region of $1.8245 \le M_{\it SW}(D)\le 1.9045$ GeV$/c^2$ around the nominal $D$ mass.
  The $D$ candidate is then combined with a negatively charged particle in the event with $p_T > 0.4$ GeV$/c$ to form a $B^{-}$ candidate. A three-dimensional kinematic fit of each decay candidate trajectory is performed by constraining the two tracks forming the $D$ candidate to a common vertex and to the nominal $D$ mass;
  the $D$ candidate and the remaining track to a separate vertex; and the reconstructed momentum of the $B^{-}$ candidate to point back to the primary \ppbar\ interaction vertex  determined from other tracks in the event.
  
The events are then divided into two non--overlapping samples, nominally classified as favored or suppressed, according to the relative charge of the $B$ candidate with the decay product of the $D$ that has been classified as the kaon.
The veto requirements applied to the $D$ mass reconstructed with the alternative particle assignment remove a large fraction of the background of favored decays from the sample classified as suppressed,
and vice versa, ensuring no overlap between the samples and a complete symmetry of the selection, which is a crucial aspect of the analysis. 
The small residual contamination of each sample from events with an incorrect identification of $D$ decay products is accounted for as part of the inclusive background $B^- \to D(\to X) \pi^-$, where $X$ are modes other than $K\pi$ (see below).
A further veto is applied to the invariant mass formed by the track from the $B$ candidate and the oppositely charged track from the $D$ candidate, again requiring it to be incompatible with the $D$ meson mass, using the same range as the first veto. This requirement suppresses the contamination from tracks from real $B$ decays that have been incorrectly labeled as $D$ decay products, and is applied symmetrically to both samples. A further suppression of this background is achieved by requiring that the transverse distance between $B$ and $D$ decay vertex is greater than 100 $\mu$m. This has the additional effect of reducing contamination from non--resonant three--body decays of the type $B^{+}\to h^{+}h^{-}h^{+}$, in which all tracks come from a common decay vertex, and where $h$ indicates either $K$ or $\pi$.  

\begin{figure*}[htbp]
\begin{center}
\includegraphics[width=3in]{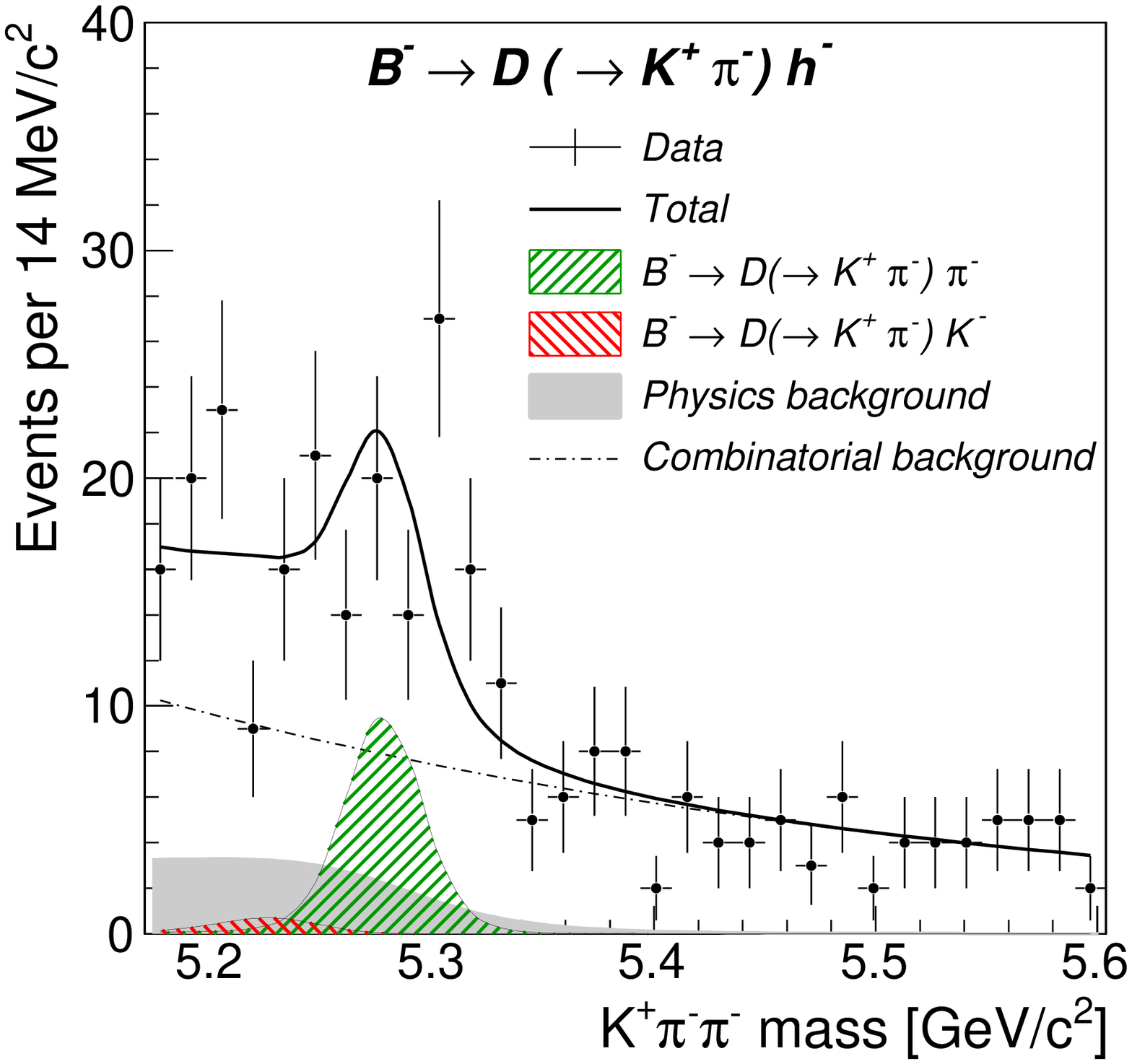}
\includegraphics[width=3in]{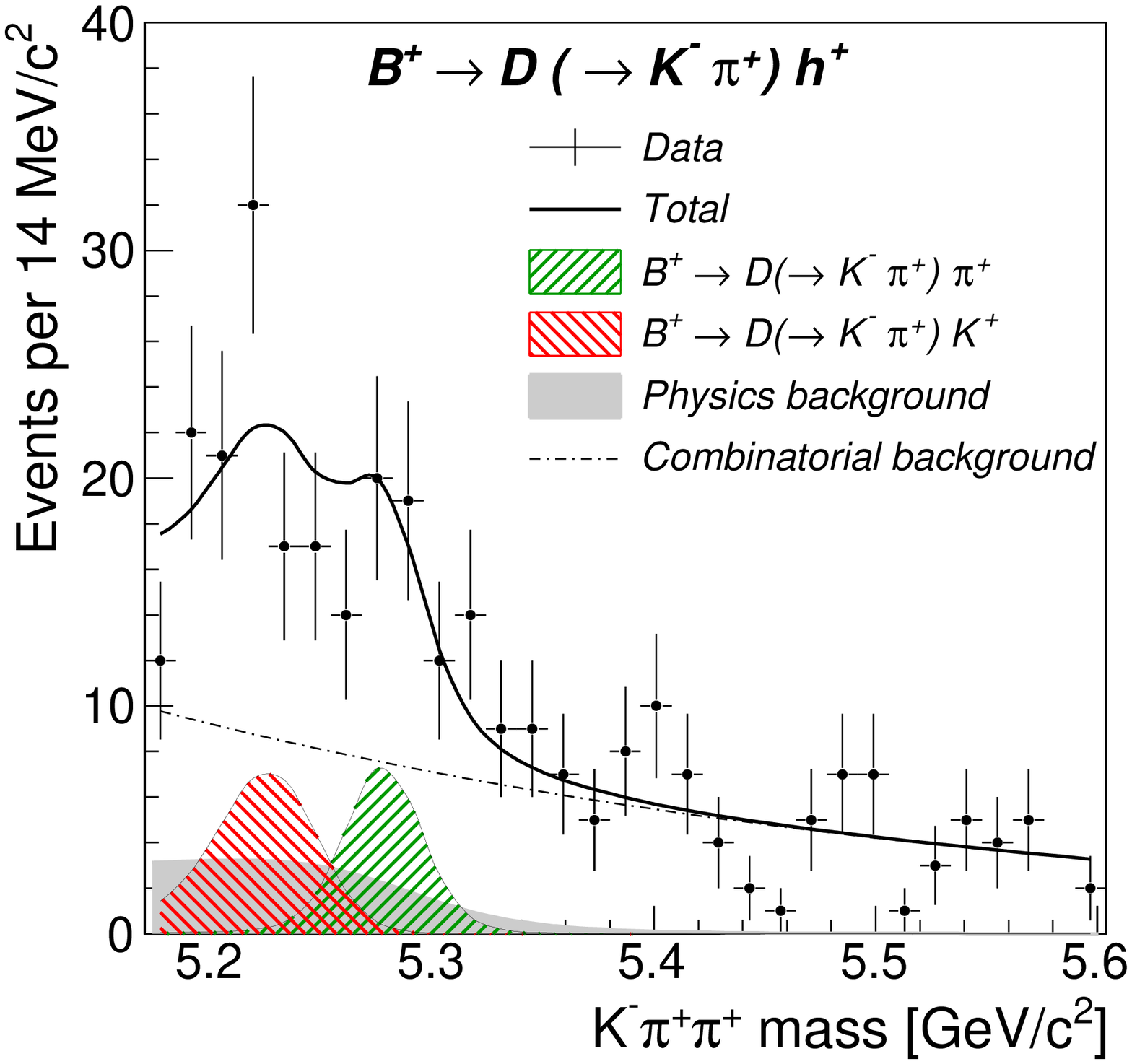}
\caption{Invariant mass distributions of $B^{\pm} \rightarrow D h^{\pm}$ for the suppressed mode (bottom meson on the left and antibottom on the right). The pion mass is assigned to the charged track from the $B$ candidate decay vertex. The projections of the common likelihood fit (see text) are overlaid. 
}
\label{mass}
\end{center}
\end{figure*}

Additional requirements are applied to the following observables: the impact parameter $d_{B}$ of the reconstructed $B$ candidate relative to the beamline; the isolation of the $B$ candidate $I_{B}$~\cite{Isolation}; the goodness of fit of the decay vertex $\chi^{2}_B$;  the significance of the $B$ hadron decay length $\sigLxy$; the angle $\alpha$ between the three-dimensional momentum of the $B$ candidate and the three-dimensional decay length; $\Delta R = \sqrt{\Delta \phi^2+\Delta\eta^2}$ between the track from the $B$ hadron and the $D$ meson; the cosine of the angle between the $D$ and the flight direction of the $B$, in the $B$ meson rest-frame, $\cos{\theta^*_D}$; the difference of the kaon probability~\cite{kaoness} values of the tracks forming the $D$ to discriminate kaon-pion pairs from pion-pion and kaon-kaon pairs, $\Delta \kappa$. 
The threshold values for all these requirements, and for the allowed $D$ mass window mentioned above, were determined by an unbiased optimization procedure, maximizing the quantity $N_S/(1.5 + \sqrt{N_B})$ \cite{articolo_punzi_sqrt}, with no use of simulated signal. The signal $N_S$ is defined as the expected rate of suppressed $B^- \to D_{\mathit{sup}} \pi^-$ events. We take advantage of our large sample of  favored $B^- \to D_{\mathit{fav}} \pi^-$ decays, using it as a model for the kinematical and particle identification properties of the suppressed decay by simply considering the swap in sign.  
The resulting requirements are the following: 
  $\sigLxy > 12$, $d_{B}< 50$~$\mu$m, $\chi^2_B<13$, $I_B ({\rm cone} = 1)>0.4$, $I_B ({\rm cone} = 0.4)>0.7$, $\alpha<0.15$, $\Delta R < 1.5$, $|\cos{\theta^*_D}|<0.6$, $\Delta \kappa>-1$.
After applying all the above selection criteria, the invariant mass of each $B^{-}\to D h^{-}$ candidate is evaluated using a nominal pion mass assignment to the particle $h^{-}$ coming from the $B$ decay. Figure~\ref{mass} shows the distributions for $B^{\pm}$ candidates.

With the help of large simulated samples of $B$ mesons, we determine that the only modes contributing non-negligible backgrounds are $B^- \to D(\to X) h^-$, $B^- \to D^{*0} \pi^-$, with $D^{*0} \to D^0 \gamma/\pi^0$, non-resonant $B^- \to K^-\pi^+\pi^-$, and $B^0 \to D_0^{*-} l^+ \nu_l$. 
The large contribution of $B^- \to D(\to K^+K^-)h^-$ reported in Ref.~\cite{ckm_belle,prd_babar} is strongly suppressed by our selection, since we reconstruct the $D$ mass in the $K\pi$ mass hypothesis.

We use an extended unbinned maximum likelihood fit, exploiting mass and particle
identification (PID) information to statistically separate the $B^{-} \to D K^{-}$ and $B^{-} \to D \pi^{-}$ signals, the combinatorial background, and the physics backgrounds. 
PID information on the track from the $B$ decay is incorporated in the kaon probability observable~\cite{kaoness}.
The extended likelihood function is defined as
$\mathcal{L} = \prod_i \mathcal{P}_{i}  \mathcal {L}_{i}$, where $i$ runs over the favored and suppressed modes, positive and negative charges. The Poisson distribution $\mathcal{P}_i$ is equal to $ \frac{\mu^{N^{\it tot}_{i}}}{N^{\it tot}_{i}  !} e^{-\mu}$, where $N^{\it tot}_{i}$ is the number of events of each sub-samples and $\mu$ is the expected mean value. The individual likelihood components have the following structure: $\mathcal{L}_{i}  =   \prod^{N^{\it tot}_i}_r  \sum_j f_{j} P_j(M_r,\kappa_r | \theta_r)$, where $f$ and $P(M_r, \kappa_r | \theta_r)$ are the fractions and the probability density functions of the signal and background modes, and $\theta_r$ are other free parameters of the fit, a mass scale parameter with respect to the nominal $B$-mass and a scale factor multiplying the width of the shapes of the $B^{-}\to D(\to K^+ \pi^-)h^-$ signals.
The fit is simultaneously performed on the favored and suppressed samples.
Common parameters are the exponential function for the combinatorial background, whose normalization and slope are determined by the fit; the functional expression for signal and background modes; and the ratio between $B^- \to D^{*0} \pi^-$ and $B^- \to D \pi^-$ fractions.
The numbers of events and the fractions of signal and background are determined by the fit and the observables are extracted from them.
We tested on simulation that our fit does not exhibit any significant bias.

The shape of the mass distribution assigned to each signal and physics background has been modeled using simulated events including the effect of final state QED radiation and parameterized with different functions. 
Systematic uncertainties are assessed by varying the values of those function parameters within their errors.

A large sample of $D^{*+}\to D^0(\to K^- \pi^+)\pi^+$ decays is used to calibrate the average \dedx\ response of the detector 
to kaons and pions, using the charge of the pion in the $D^{*+}$ decay to determine the identity of the $D$ decay products. 
The shape of the $\kappa$ distribution is calibrated within our own sample, by using  kaons and pions from the decay of the $D$ meson in the favored sample.
Uncertainties on the calibration parameters are included in the final systematic uncertainty of $A$, $R$ and $R^{\pm}$, taking into account the full correlation matrix of the parameters characterizing the shape of the $\kappa$ distribution\cite{dott_paola}. 

The $B^{-} \rightarrow D K^{-}$ and $B^{-}\rightarrow D\pi^{-}$ event yields obtained from the fit to the data are reported in Table \ref{fit_res}. 
Fit projections on the invariant mass distributions are given in Fig.~\ref{mass}. They provide a consistent description of the observed distributions in the data.
We find evidence for a signal in the $B^- \rightarrow D K^-$ suppressed mode with a significance of 3.2$\sigma$. 
The significance is evaluated by comparing the likelihood-ratio observed in data with the distribution expected in statistical trials. Several distributions are generated corresponding to different choices of systematic parameters. The quoted significance corresponds to the distribution yielding the most conservative $p$-value.
 \begin{table*}[htb]
 \caption{$B^{-} \rightarrow D K^{-}$ and $B^{-}\rightarrow D \pi^{-}$
 event yields obtained from the fit to the data. Only statistical uncertainties are quoted.}
  \begin{center}
    \begin{tabular}{c c c c c}
      \hline\hline
      \footnotesize{$D$ mode} & \footnotesize{$B^+\rightarrow D \pi^+$} &
      \footnotesize{$B^-\rightarrow D \pi^-$} & \footnotesize{$B^+\rightarrow D
      K^+$} & \footnotesize{$B^-\rightarrow D K^-$}\\
      \hline
      $K^-\pi^+$ (favored) & $9882 \pm 103$ & $9892 \pm 103$ & $694 \pm 39$ & $767 \pm 41$\\
      $K^+\pi^-$ (suppressed) & $24 \pm 9$ & $31 \pm 10$ & $29 \pm 9$ & $3 \pm 8$\\
      \hline\hline
    \end{tabular}
    \label{fit_res}
  \end{center}
\end{table*}
The raw fit results are then corrected for the reconstruction efficiency $\epsilon$, due to different probabilities of $K^+$, $K^-$, $\pi^+$ and $\pi^-$ to interact with the tracker material.
We use previous measurements of $\frac{\epsilon(K^+)}{\epsilon(K^-)}=1.0178\pm 0.0023 (\rm{stat}) \pm 0.0045 (\rm{syst})$
and $ \frac{\epsilon(\pi^+)}{\epsilon(\pi^-)}=0.997\pm 0.003 (\rm{stat})\pm 0.006 (\rm{syst})$~\cite{acp_corr}.
We extract $\frac{\epsilon (K^- \pi^+)}{\epsilon (K^+ \pi^-)} = 0.998 \pm 0.015(\rm{stat}) \pm 0.016(\rm{syst})$ from our own sample of favored $B^- \to D \pi^-$ decays.
\begin{table*}[htb]
 \caption{Summary of systematic uncertainties.}
  \begin{center}
    \begin{tabular}{l c c c c c c c c}
       \hline\hline
            Source & $R(\pi)$ & $R^+(\pi)$ & $R^-(\pi)$ &$R(K)$ & $R^+(K)$ &$R^-(K)$ &$A(\pi)$ & $A(K)$\\
      \hline
      \dedx\ model & $<$0.0001 & $<$0.0001 & $<$0.0001 & 0.0001 & 0.0003 & 0.0001 & $<$0.01 & $<$0.01\\
      $B^- \to D(\to X)\pi^-$ shape & 0.0004 & 0.0004 & 0.0004 & 0.0025 & 0.0026 &0.0026 &0.01 & 0.09\\
	Other backgrounds & $<$0.0001 & $<$0.0001 & $<$0.0001 & 0.0006 & 0.0006 &0.0005 &$<$0.01 & 0.02\\
      Efficiency & $<$0.0001 & $<$0.0001 & $<$0.0001 &  0.0003 & 0.0009 &0.0001 & 0.01 &  $<$0.01\\
       \hline
      Total  & 0.0004 & 0.0004 &0.0004 & 0.0026 & 0.0028 &0.0027 & 0.02 & 0.09\\
      \hline\hline
    \end{tabular}
    \label{systematics}
  \end{center}
\end{table*}

Systematic uncertainties are determined by repeating the fit changing the mass and the $dE/dx$ model (Table~\ref{systematics}). 
The dominant contribution is the uncertainty on the $B^- \to D(\to X)\pi^-$ shape. This is the largest physics background, and it lies under the signal peak.

In summary, we find evidence for the $B^- \rightarrow D(\to K^+ \pi^-) K^-$ suppressed mode with a significance of 3.2 Gaussian standard deviations.
We measure the ratios of the suppressed $([K^+\pi^-]_D K^-/\pi^-)$ to favored $([K^-\pi^+]_D K^-/\pi^-)$ branching fractions 
 $R(K) = [22.0\pm 8.6(\rm{stat})\pm 2.6(\rm{syst})]\times 10^{-3}$, 
$R^+ (K)   =  [42.6\pm 13.7 (\rm{stat})\pm 2.8 (\rm{syst})]\times 10^{-3}$,
$R^- (K)   =  [3.8\pm 10.3 (\rm{stat})\pm 2.7 (\rm{syst})]\times 10^{-3}$
  and 
 $R(\pi) = [2.8\pm 0.7(\rm{stat})\pm 0.4(\rm{syst})]\times 10^{-3}$,  
$ R^+ (\pi)   =  [2.4\pm 1.0 (\rm{stat})\pm 0.4 (\rm{syst})]\times10^{-3}$,
$R^- (\pi)   =  [3.1\pm 1.1 (\rm{stat})\pm 0.4 (\rm{syst})]\times 10^{-3}$
 as well as the direct  $\mathit{CP}$-violating asymmetries
 \begin{eqnarray}
  A(K) &  = & -0.82\pm 0.44(\rm{stat})\pm 0.09(\rm{syst}), \nonumber\\
  A(\pi) & = &  0.13\pm 0.25(\rm{stat})\pm 0.02(\rm{syst})\nonumber.
 \end{eqnarray}
The observed asymmetry $A(K)$ 
deviates from zero by 2.2 standard deviations.
  
  These measurements, performed here for the first time in 
hadron collisions, are in agreement with previous measurements from 
\babar\ \cite{prd_babar} and Belle \cite{ckm_belle} with comparable uncertainties.
  These results can be combined with other $B^{-} 
\rightarrow D K^{-}$ measurements to improve the  
determination of the CKM angle $\gamma$. 

\section*{Acknowledgments}
We thank the Fermilab staff and the technical staffs of the participating institutions for their vital contributions. This work was supported by the U.S. Department of Energy and National Science Foundation; the Italian Istituto Nazionale di Fisica Nucleare; the Ministry of Education, Culture, Sports, Science and Technology of Japan; the Natural Sciences and Engineering Research Council of Canada; the National Science Council of the Republic of China; the Swiss National Science Foundation; the A.P. Sloan Foundation; the Bundesministerium f\"ur Bildung und Forschung, Germany; the Korean World Class University Program, the National Research Foundation of Korea; the Science and Technology Facilities Council and the Royal Society, UK; the Institut National de Physique Nucleaire et Physique des Particules/CNRS; the Russian Foundation for Basic Research; the Ministerio de Ciencia e Innovaci\'{o}n, and Programa Consolider-Ingenio 2010, Spain; the Slovak R\&D Agency; the Academy of Finland; and the Australian Research Council (ARC).

 \bibliography{bibliografia}

\begin{thebibliography}{31}
\bibitem[{charge()}]{charge_conjugate}
{\bibinfo{title}{{$D$ indicates $D^0$ and $\bar{D}^0$, and the charge conjugate state is implied throughout the paper, except in formulas and sentences where both are mentioned explicitly.}}}
\bibitem[{rb()}]{rb_def}
{\bibinfo{title}{{$r_B$ is defined as the magnitude of the amplitude ratio of the suppressed process $b \to u$ over the favored process $b \to c$, $r_B = \left| \frac{\mathcal{M}(b\to u)}{\mathcal{M}(b\to c)} \right| $. Since the suppressed transition is associated to the $B^- \to \bar{D}^0 K^-$ decay and the favored to $B^- \to D^0 K^-$, $r_B$ corresponds also to $\left| \frac{\mathcal{M}(B^- \to \bar{D}^0 K^-)}{\mathcal{M}(B^- \to {D}^0 K^-)} \right|$. In the text we will distinguish between $r_B$ of the kaon, $r_B(K)$, and of the pion, $r_B(\pi)$. The definitions for the pion are analogous to the definitions for the kaon.}}}
\bibitem[{CDF()}]{CDF-coordinates}
  {\bibinfo{title}{{CDF II uses a cylindrical coordinate system in which
        $\phi$ is the azimuthal angle, $r$ is the radius from the nominal beam line,
        and $z$ points in the proton-beam direction, with the origin at the center of
        the detector. The transverse plane is the plane perpendicular to the $z$
        axis.}}}
\bibitem[{Iso()}]{Isolation}
  {\bibinfo{title}{{Isolation is defined as
        $I_{B}=p_T(B)/(p_T(B)+\sum_ip_{Ti})$, where $p_T(B)$ is the transverse
        momentum of the $B$ candidate, and the sum runs over all other tracks within
        a cone in the $\eta-\phi$ space around the $B$ flight-direction. 
        Its value is typically higher for bottom--flavored hadrons than for random track combinations.}}}
\end{thebibliography}
 
 \end{document}